# Modeling edge effects in Graphene Nanoribbon Field-effect Transistors with real and mode space methods

Pei Zhao and Jing Guo

*Department of Electrical and Computer Engineering, University of Florida, Gainesville, Florida 32611*

**Abstract**

A computationally efficient mode space simulation method for atomistic simulation of a graphene nanoribbon field-effect transistor in the ballistic limits is developed. The proposed simulation scheme, which solves the nonequilibrium Green's function coupled with a three dimensional Poisson equation, is based on the atomistic Hamiltonian in a decoupled mode space. The mode space approach, which only treats a few modes (subbands), significantly reduces the simulation time. Additionally, the edge bond relaxation and the third nearest neighbor effects are also included in the quantum transport solver. Simulation examples show that, mode space approach can significantly decrease the simulation cost by about an order of magnitude, yet the results are still accurate. This article also demonstrates that the effects of edge bond relaxation and third nearest neighbor significantly influence the transistor's performance and are necessary to be included in the modeling.

## I. Introduction

The motivation to extend the Moore's Law drives strong interest for searching new transistor channel materials beyond silicon. Carbon related materials such as carbon nanotube and graphene show promising for application in nanoelectronic system. Comparing with the carbon nanotube, the graphene,[1,2] a two dimensional lattice cutting from the Graphite, shows potential for integration with planar fabrication process.[3,4] Although the metallic property of 2D graphene sheet limits its application for semiconducting device, graphenen nanoribbon (GNR),[5] a narrow ribbon cutting from graphene, is a quasi-1D material with a bandgap,[6] which depends on the width of the nanoribbon and their crystallographic direction. A recent experiment demonstrated that all sub-10nm-wide GNRs are semiconducting,[7] which makes them more attractive for electronic device applications.

The numeral simulation for graphene nanoribbon field-effect transistors (GNRFETs) can be achieved by self consistent solving the nonequilibrium Green's function (NEGF) equation[8] coupled to a three dimensional Poisson equation in ballistic limits with a real space basis. The real space approach is an atomistic simulation of small electronic device, which also explore and understand the effect of contact, interfaces, and defects.[9-12] Although the real space approach demonstrates rigorous results, even for the nano-scale device the associated computational burden is heavy. Additionally, strong confinement in the width direction causes the separation between the subbands to be large in the channel of a GNRFET. The real space representation, however, can only compute all modes together, and cannot elucidate effects of each subband. A mode space approach, which solves the Green's function in a mode space representation, can significantly reduce the computational cost and examine the effects of each subbands (modes), as shown for silicon FETs[13] and carbon nanotube (CNT) FETs.[14] Significant differences, however, exist between CNTs and GNRs due to the existence of edges in GNRs, which makes it necessary to identify a different mode space basis set for the GNRFET and to treat the edge effects, such as the edge bond relaxation and the third nearest neighbor coupling.[15-17]

Our purpose in this paper is to introduce an atomistic mode space approach into the NEGF simulator to simulate the same device geometry as the real space approach for GNRFETs. The mode space approach decouples the 2D GNR lattice into 1D lattice, which greatly saves the computational cost by about an order of magnitude. Based on carefully validated approximations, the close agreement can be achieved between the results of real space approach and results of mode space approach. Furthermore, the edge bond relaxation and third nearest neighbor effects, which were demonstrated to influence the GNR band structure, but do not exist in the CNTs are modeled in the real and mode space approaches. Simulation examples show these effects significantly affect the transistor characteristics.

## II. NEGF Treatment of Quantum Transport in GNRFETs

### A. Real space approach

This section summarizes the NEGF treatment of a GNRFET in a real space basis for the sake of completeness and as the starting point for the discussion of the mode space approach. The real space basis set is using the simple $\pi$-orbital nearest neighbor tight binding model. A semiconducting armchair-edge GNR (A-GNR), whose bandgap mostly stems from the quantum confinement in the width direction, is modeled as the transistor channel. As an example, we model an $n$=12 A-GNR, in which the index $n$ denotes the number of dimmer carbon atom lines along the transport direction and is proportional to the GNR width.[18] The first step is to write down the Hamiltonian matrix for the isolated channel. As shown in Fig. 1 the A-GNR lattice, similar to a carbon nanotube lattice, is also composed of A and B sublattices, but edges exist in a GNR lattice. The A and B sublattices alternately couple to each other in x-direction. If the number of carbon atoms in the entire channel is $N$, the size of the Hamiltonian matrix is $N \times N$:

$$H = \begin{bmatrix} \alpha_1 & \beta_2^+ & & & & \\ \beta_2^+ & \alpha_2 & \beta_1 & & & \\ & \beta_1 & \alpha_3 & \beta_2^+ & & \\ & & \beta_2^+ & \alpha_4 & \beta_1 & \\ & & & \beta_1 & \alpha_5 & \cdots \\ & & & & \cdots & \cdots \end{bmatrix} \quad (1)$$

where the $(n/2) \times (n/2)$ submatrix $[\boldsymbol{\alpha_i}]$ describes coupling within the A or B atom lines in width direction, and $(n/2) \times (n/2)$ submatrix $[\boldsymbol{\beta}]$ describes the coupling between adjacent atom lines. Due to the nearest neighbor tight binding approximation, carbon atoms within each atom line are uncoupled to each other so that $[\boldsymbol{\alpha_i}]$ is a diagonal matrix. The value of each diagonal entry is the electrostatic potential at that atom site. The matrices $[\boldsymbol{\beta}]$, which describe the coupling between adjacent atom lines, have two types of coupling matrices $[\boldsymbol{\beta_1}]$ and $[\boldsymbol{\beta_2}]$ as shown in Fig. 1. The detailed description of the coupling matrices for CNTs can be checked in.[14] For GNRs coupling matrix $[\boldsymbol{\beta_1}] = \boldsymbol{t}[\boldsymbol{I}]$ is the same with CNTs'. However GNRs, with edges at two sides, no periodic condition exists, the top right element of coupling matrix $[\boldsymbol{\beta_2}]$ should be zero:

$$[\beta_2]=t_0\begin{bmatrix}1 & & \cdots & 0\\ 1 & 1 & & \\ & 1 & 1 & \cdots\\ & & \cdots & \cdots\end{bmatrix} \quad (2)$$

Having specified the Hamiltonian matrix for the channel, the next step is to compute the self-energy matrix $\boldsymbol{\Sigma_S}$ and $\boldsymbol{\Sigma_D}$ for source and drain region. The recursive relation[13] relates the surface Green's functions is

$$g_1=[(E+i0^+)I-\alpha_1-\beta_2^+g_2\beta_2]^{-1}$$

$$g_2=[(E+i0^+)I-\alpha_2-\beta_1^+g_3\beta_1]^{-1} \quad (3)$$

where matrix $\boldsymbol{g_i}$ is the surface Green's function for the $i$th atom line in the source region. And $\boldsymbol{g_1}=\boldsymbol{g_3}$ due to the periodicity of the A and B sublattices. From above matrix equations with two unknown matrices $\boldsymbol{g_1}$ and $\boldsymbol{g_2}$, the surface Green's function can be solved. Non-zero submatrix of source self energy matrix is $\boldsymbol{\Sigma_S^{1,1}=\beta_1 g_1 \beta_1^+}$. Similar approach can be used to solve the drain self energy matrix.

The retarded Green's function is then determined by

$$G(E)=[(E+i0^+)I-H-\Sigma_S-\Sigma_D]^{-1} \quad (4)$$

which describes how the graphene nanoribbon channel connect to the two contacts by the self energy matrices. The local density of states resulting from the source/drain injected states is calculated using

$$D_{S(D)}=G\Gamma_{S(D)}G^+ \quad (5)$$

where $\boldsymbol{\Gamma_{S(D)}=i(\Sigma_{S(D)}-\Sigma_{S(D)}^+)}$ is the energy level broadening due to the source/drain contacts. The charge density is calculated by

$$Ne=\int_{-\infty}^{+\infty}dE\cdot\mathrm{sgn}[E-E_N]\{D_S(E)f(\mathrm{sgn}[E-E_N](E-E_{FS}))+D_D(E)f(\mathrm{sgn}[E-E_N](E-E_{FD}))\} \quad (6)$$

where $E_{FS}$, $E_{FD}$ is the source and drain Fermi level. Charge density is then fed back to the Poisson equation[9-11] for self consistent solutions. Once self consistent is achieved, the source-drain current is computed from

$$I = \frac{2e}{h}\int dE \cdot T(E)[f(E-E_{FS}) - f(E-E_{FD})] \quad (7)$$

where $T(E)=Trace(\Gamma_1 G \Gamma_2 G^+)$ is the source/drain transmission.

### B. Mode space approach

In this simulation scheme, the Green's function is solved in a mode space representation. The most important step is to determine the basis transform matrix. Because graphene nanoribbon only has one atom layer, the channel material is a 2D problem. The mode space approach decouples the 2D lattice into several 1D lattices by a basis transform in the width direction of the GNR. Key to this problem is to identify the new basis functions (the modes) in the width direction. For a carbon nanotube, the modes must satisfy the periodic boundary condition in the circumferential direction.[14] However, for a graphene nanoribbon, the "particle-in-a-box" boundary condition is imposed at the GNR edges, and we identified the following mode space basis set for the A type and B type atom lines along the width direction, as shown in Fig. 1(a),

$$\phi_{iv} = \frac{2}{\sqrt{n+1}}\sin([\frac{v\pi}{n+1}2i]) \quad \text{(A sublattice)}$$

$$\varphi_{iv} = \frac{2}{\sqrt{n+1}}\sin([\frac{v\pi}{n+1}(2i-1)]) \quad \text{(B sublattice)} \quad (8)$$

where $i$=1,2…($n$/2) is the index for atoms in real space atom lines in width direction, $v$=1,2…($n$/2) is the index for modes in mode space representation. A (B) is two kinds of sublattices. The solutions are satisfied with the quantum well boundary condition in width direction, thus give us the basis transform matrix **[*V*]**. Then we perform this basis transformation on the two dimensional graphene nanoribbon lattices to decouple the problem in *n/2* one dimensional mode space lattices

$$\alpha_i^{'} = [V]^{+} \alpha_i [V]$$

$$\beta_1^{'} = [V]^{+} \beta_1 [V]$$

$$\beta_2^{'} = [V]^{+} \beta_2 [V] \quad (9)$$

where $\boldsymbol{\alpha_i'}$, $\boldsymbol{\beta_1'}$, and $\boldsymbol{\beta_2'}$ are all diagonal matrices. There are no matrix elements between different modes in the width direction after the basis transformation, which means that no interaction exists between modes. If we reorder the basis according to the modes basis, the Hamiltonian matrix is

$$[H] = \begin{bmatrix} H_1 & & & & \\ & H_2 & & & \\ & & \ddots & & \\ & & & H_q & \\ & & & & \ddots \end{bmatrix} \quad (10)$$

where the $\boldsymbol{H_q}$ is the Hamiltonian matrix for the *q*th mode

$$[H_q] = \begin{bmatrix} U_1 & b_{2q} & & \\ b_{2q} & U_2 & b_{1q} & \\ & b_{1q} & U_3 & b_{2q} \\ & & \cdots & \cdots \end{bmatrix} \quad (11)$$

where $U_i$, which is the diagonal element in matrix $\boldsymbol{\alpha_i'}$, is the electrostatic potential at the *i*th atom in *q*th mode. $b_{1q}=t_0$ and $b_{2q}=2t_0\cos(q\pi/(n+1))$, the *q*th diagonal element of $\boldsymbol{\beta_1'}$, and $\boldsymbol{\beta_2'}$, are the coupling parameters between nearest neighbor in *q*th mode 1D lattice. The mode space decoupled lattices is shown in Fig. 1(b). After the basis transform, each element of Hamiltonian matrix in mode space is only a number not an $n \times n$ submatrix, thus the size of the problem significantly reduced.

The self energy calculation in mode space uses the same approach as we introduced in real space. However, for mode space $g_i$, $\beta_1$, and $\beta_2$ are all number rather than matrices. Thus equation (3) can be analytically solved

$$g_{1q} = \frac{(E-U_1)^2 + b_{1q}^2 - b_{2q}^2 + \sqrt{[(E-U_1)^2 + b_{1q}^2 - b_{2q}^2]^2 - 4(E-U_1)^2 b_{1q}^2}}{2b_{1q}^2(E-U_1)} \tag{12}$$

The source self energy for the $q$th mode is $\Sigma_{Sq}=(b_{1q})^2 g_{1q}$. Similar expression can be found for drain contact with the different electrostatic potential at the drain end. When Hamiltonian matrix and self energy matrices are determined, the Green's function has the same form as equation (4). Then the calculation of carrier density and current also follow the same scheme in real space.

**C. Effects of Edge bond relaxation**

Previous real and mode space approach are both based on the π-orbital nearest neighbor tight binding approximation, which assumes the same TB parameter for all bonds.[14] The existence of edges in GNRs, however, makes the effect of edge bond relaxation and the third nearest neighbor coupling pronounced. The edge effects impose new challenges for GNRFETs simulation beyond the CNTFETs simulation, which are addressed in this section and the next section.

The edge bond relaxation can be implemented into the Hamiltonian matrix with an extra matrix

$$[\Delta\beta_1] = \begin{bmatrix} \Delta t_0 & & \\ & 0 & \\ & & \ddots \end{bmatrix} or \begin{bmatrix} \ddots & & \\ & 0 & \\ & & \Delta t_0 \end{bmatrix} \quad (13)$$

where the $\Delta t_0$ is the additional part for the carbon-to-carbon nearest neighbor coupling at edge sides. In our example 12-AGNR, the edge bond relaxation can only exist at one side of the GNR shown in Fig. 1(a), thus we have two kinds of the matrix forms. Then $\boldsymbol{\beta_{1(edge)}}= \boldsymbol{\beta_{1(simple)}}+\boldsymbol{\Delta\beta_1}$, where $\boldsymbol{\beta_{1(simple)}}$ is the coupling matrix for the nearest neighbor tight bonding model, $\boldsymbol{\beta_{1(edge)}}$ is the coupling matrix including the edge bond relaxation.

For real space scheme, including the edge bond relaxation only changes some elements of the Hamiltonian matrix, the simulation process still follow the same equation. Edge bond relaxation makes the matrix a little more complex, thus more simulation time consumption is required. For mode space, basis transform matrix, as we derived before, can still be used to transform the new Hamiltonian matrix:

$$\beta'_{1(edge)} = [V]^+ \beta_{1(edge)} [V]$$

$$\beta'_{2(edge)} = [V]^+ \beta_{2(edge)} [V] \quad (14)$$

Because edge effect does not influence coupling matrix $\boldsymbol{\beta_2}$, thus $b_{2q} =2t_0cos(q\pi/(n+1))$ does not change. Although the edge bond relaxation only exists at edge side, when performance the basis transform, each mode will be affect by the edge effect $\Delta t_0$. An analytical expression can be derived from the matrix equation (14) and obtain the $q$th diagonal entry of $\boldsymbol{\beta_1'}_{(edge)}$: $b_{1q} =t_0+4\Delta t_0 sin^2(q\pi/(n+1))/(n+1)$. The basis transform of $\boldsymbol{\Delta\beta_1}$ also results in small coupling between different modes (small off diagonal elements), which is neglected as an approximation. The exactly same $E$-$k$ relation as the real space Hamiltonian to the first order of $\Delta t_0/ t_0$ can be obtained from the mode space Hamiltonian, [17] and the accuracy of this approximation on the GNRFET will be further examined by numerical simulations later.

For calculation of the contact self energy, equation (12) can still be used for self energy calculation in mode space. The only difference is $b_{1q}$ is changing due to the edge bond relaxation.

**D. Effect of third nearest neighbor**

However, recent study[16,17] points that edge bond relaxation could only explain part of the mismatch, interaction across the hexagon should also be included as second and third nearest neighbor coupling, as solid lines across the hexagon in Fig. 1(a). Second nearest neighbor, which only shift the dispersion relation in the energy direction but not change the band structure, can be ignored.[16,17] Third nearest neighbor interaction, although much weaker than the first nearest neighbor interaction, is necessary to be considered in our model.

To implement the third nearest neighbor effect into the real space, we only need to add the third nearest neighbor interaction $t_3$ into coupling matrix [$\boldsymbol{\beta_1}$], coupling matrix [$\boldsymbol{\beta_2}$] still keep the same from:

$$[\beta_{1(3NN)}] = \begin{bmatrix} t_0 & t_3 & & & & \\ t_3 & t_0 & t_3 & & & \\ & t_3 & t_0 & t_3 & & \\ & & t_3 & t_0 & t_3 & \\ & & & t_3 & t_0 & t_3 \\ & & & & t_3 & t_0 \end{bmatrix} \tag{15}$$

For real space the simulation can still be directly achieved. However when including the interaction beyond nearest neighbor, the Green's function will be not sparse as before, which leads around the twice the total simulation time increasing.

For mode space, the third nearest neighbor effect need to be treated separately: interaction between different modes and interaction within same mode. For the first case, third nearest neighbor interaction between modes including in coupling matrix $\boldsymbol{\beta}_{1(3NN)}$ as equation (15). Basis transform could be performance on the new matrix:

$$\beta'_{1(3NN)} = [V]^{+} \beta_{1(3NN)} [V] \tag{16}$$

diagonal elements of $\boldsymbol{\beta}'_{1(3NN)}$ give the new coupling parameter $b_{1q}$ in mode space. The second case, interaction in transport direction, makes Hamiltonian matrix not a tri-diagonal any longer:

$$H_q = \begin{bmatrix} U_1 & b_{2q} & 0 & t_3 & & \\ b_{2q} & U_2 & b_{1q} & 0 & & \\ 0 & b_{1q} & U_3 & b_{2q} & 0 & t_3 \\ t_3 & 0 & b_{2q} & U_4 & b_{1q} & 0 \\ & & 0 & b_{1q} & U_5 & \cdots \\ & & t_3 & 0 & \cdots & \cdots \end{bmatrix} \tag{17}$$

where $t_3$ describes the third nearest neighbor coupling in the transport direction, which set to $t_3$=0.2eV in our example. $b_{1q}$ is the $q$th diagonal entry in $\boldsymbol{\beta}_1'_{(3NN)}$. Similar to what we discussed in edge bond relaxation, after the basis transform, coupling matrix $\boldsymbol{\beta}_1'_{(3NN)}$ in mode space is also not perfect diagonal. Similar to the edge bond relaxation effect, neglecting the small off-diagonal elements is accurate to the first order of $t_3/\ t_0$. The accuracy of this approximation will also be further checked in section 3.

Including third nearest neighbor will also affect the self energy calculation in mode space, as shown in Fig. 2, at the source end of the channel two carbon atoms will couple with the contact. Because the coupling of the mode space lattice is beyond the nearest neighbor as shown in Fig. 2, the simple analytical expression, equation (12) cannot be applied. If we group A and B two atoms together as a new cell, the coupling between adjacent new cell can be described by a $2 \times 2$ matrix, and each new cell only couples to its nearest neighbors. Thus the recursive relation for the surface green's function, equation (3), can be utilized to numerically calculate the contact self energy, with each quantity being a $2 \times 2$ matrix.

Using the iterative loop to solve the self energy matrix will increase the mode space simulation time. An improved way is using Sancho-Rubio iterative method,[19] which reduces the total simulation time five times shorter in our simulation example. The matrix equation in mode space is much smaller than that in real space, thus the computational cost is much less expensive.

## III. Results and Discussions

According to the edge shape, graphene nanoribbon has zigzag GNR and armchair GNR two different with distinguished property. For zigzag GNR the bandgap collapses at a finite source-drain bias,[20] and is not suitable to be used as the MOSFET-type device. In this work, therefore, we focus on exploring the physical properties and device performance of armchair GNRFETs.

To explore the performance of the armchair GNRFETs as shown in Fig. 3, a double gate as is used. The GNR is placed between two insulator layers, assumed to be $SiO_2$ of 1.5nm thickness. The source/drain region is assumed to be doped GNR with $5x10^{-3}$ dopant / atoms. The channel is intrinsic and the gate length equals the channel length as the 10nm. The source/drain is assumed as the physical extension of the channel (having the same width). This 1D contact is easier to deal with compared with 2D contact.[11] In the simulation, a tight-binding parameter of $t_0$ = 2.7 eV is used,[15] and edge bond relaxation introduced with coefficient $C_{edge}$=1.12.[15]

Using the NEGF approach, the local density of states versus energy and position is calculated in Fig. 4 at on state ($V_G$=$V_D$=0.5V). Fig. 4(a) is the result from real space calculation. The solid lines correspond to the first conduction band edge and first valence band edge. In the conduction band region, the first and second subbands are clearly visible by comparing the real space results to the mode space results in Fig. 4(b) and (c). The oscillation patterns are due to the quantum mechanical reflections. The separations of the first and second subbands are around 0.4eV which is close to the analytical results obtained by dispersion relationship. A quantum well can also be seen near the top of the valence band barrier. In fact, the second subband contributes little in the carrier transport, which can be examined in the mode space approach. Since the first subband is much more dominant than other subbands, then only simulating the first subband using mode space approach is more effective than real space approach, which calculating all the modes.

Next, we plot the transmission probability versus the energy at on state for three models in Fig. 5. Model 1 (solid line) is the simple tight bonding model without edge bond relaxation and the third nearest neighbor interaction. Model 2 (dashed line) considers edge bond relaxation only. Model 3 (dotted line) considers both effects. In the middle energy region as the forbidden state, no states can be occupied thus transmission is zero. At low energy in conduction band region, electrons need to overcome the potential barrier, thus the transmission probability gradually increasing. At high energy level, electron can directly transport from source to drain, the transmission probability is always one. For three models, the transmission probability share the similar shape of curve, the only differences is bandgap region becoming smaller when including edge bond relaxation and third nearest neighbor at the case n=12, which is agree with calculation results in.[17]

Fig. 6 and Fig. 7 plot the potential profile and the carrier density along the carrier transport direction at on-state. Fig. 6 is for model 1. The mode space simulation (crosses) excellently reproduces the results of the real space approach (solid line). The good agreement is based on the approximation that in width direction the potential in graphene nanoribbon is uniform. Fig. 7(a),(b) are the results for model 2 and Fig. 7(c),(d) are corresponding to model 3. When including the edge bond relaxation and third nearest neighbor results of mode space approach (dashed lines) cannot perfect catch the results of real space approach (solid lines). However two curves still almost overlap with each other. The small mismatch can be attributed to the approximation made in basis transform of the Hamiltonian in mode space: small interaction between modes (off-diagonal elements) is ignored.

Fig. 8 compares the $I_{DS}$-$V_{DS}$ characteristics ($V_G$=0.5V) of the real space and mode space approaches for three models. The first step, aiming to examine how edge bond relaxation and third nearest contribute to the device performance, only compare real space results (solid lines) for these three models. Under the same bias, including the edge bond relaxation and third nearest neighbor the currents increase corresponding to the bandgap decreasing. Including edge bond relaxation the on current will increase 1.7 times, and with both edge bond relaxation and third nearest the increment will be 3 times which shows that these two effects will significant influence the transistor's performance. Thus edge bond relaxation and third nearest neighbor are the crucial effects when simulating the GNRFETs characteristics. Table 1 shows that without treating the edge effect (model 1), the mode space method agrees with the real space calculation within 1% in terms of current. Treatment of the edge effects in the mode space requires further approximations as described in section 2.3 and 2.4, and results in a larger error. But the accuracy is still within around 5%. Table 1 also shows that the mode space approach reduces the computational cost by about one order of magnitude (the simulations were run on a single 3.0GHz CPU).

## IV. Summary

In this work, we describe a decoupled mode space approach based on the NEGF formalism coupled with a three dimensional Poisson equation, which could be used to simulate the GNRFETs. A new basis transform matrix, which is different from CNTFET and MOSFET, is derived for a graphene nanoribbon. The modeled device is a DG MOSFET like GNRFET, the channel is assumed to be armchair graphene nanoribbon, similar simulation can also been achieved in real space basis, which requests expensive computation cost. Mode space approach can greatly reduce the simulation cost by about one order of magnitude in our examples, yet still accurate enough. Furthermore, the edge bond relaxation and third nearest neighbor, which not pronounced have important effect in CNT, significantly influence characteristics of GNRFETs.

Acknowledgements: This work was supported in part by ONR N000140810861, NSF ECCS-0846543, NSF ECCS-0824157.

FIGURE CAPTIONS

Fig. 1 (a) The schematic diagram of an $n$=12 Armchair GNR. The circles are the A-type carbon sublattice, and the triangles are the B-type carbon sublattice. The coordinate system is also shown: $z$ is the width direction, and $x$ is the carrier transport direction. Edge bond relaxation (dashed lines) is shown. Solid lines across the hexagon shows the second (2NN) and third nearest neighbor (3NN) coupling. (b) The mode space decoupled 1D lattice, each lattice is cut from the solid square in (a). A basis transform transforms the real space 2D lattice into 1D problem. Carbon atoms in z direction have no interaction.

Fig. 2 Computing the contact self-energy for the GNRFET in the mode space. The third nearest coupling ($b_3$) is included. A (circles) and B (triangles) sublattices atom can be grouped together to make the coupling only exists between the nearest neighbor.

Fig. 3 The modeled double-gated GNRFETs with heavily-doped (the doping density of the source/drain extension is 5 x $10^{-3}$ dopant/atom). The channel is intrinsic and the gate length is equal to the channel length as 10nm. The oxide thickness is 1.5nm.

Fig. 4 (a) Local density of states (LDOS) computed by the real space approach and (b) Local density of states (LDOS) of the first and second subbands computed by the mode space approach. ($V_G$=0.5V and $V_D$=0.5V) The lines are the conduction band edge and valence band edge.

Fig. 5 Transmission vs. Energy computed by the real space approach ($V_G$=$V_D$=0.5V), for three models: model 1 (solid line) not including edge bond relaxation and third nearest neighbor, model 2 including the edge bond relaxation (dashed line) and model 3 including both effects (dotted line). The decreasing of the bandgap could be observed when including the edge bond relaxation and third nearest neighbor.

Fig. 6 (a) Potential profile computed by the real space approach (solid line) and mode space approach (crosses) at $V_G$=0.5V, $V_D$=0.5V. (b) The charge density computed by the real space approach (solid line) and mode space approach (crosses) at the same bias. All results are not including the edge bond relaxation and third nearest neighbor (model 1).

Fig. 7 (a),(b) Potential profile and charge density of model 2 computed by the real space approach (solid line) and mode space approach (dashed line) at $V_G$=0.5V, $V_D$=0.5V. For including the edge bond relaxation, real and mode space approach calculation results not perfect match with each other. (c),(d) Potential profile and charge density of model 3 computed by the real space approach (solid line) and mode space approach (dashed line) at the same bias. Further including the third nearest neighbor, the mismatch become more clearly, however results is still very close.

Fig. 8 $I_D$ vs. $V_D$ characteristics for the model device from real (lines) and mode-space (circles) solution at $V_G$=0.5V. Close agreement between two approaches can be achieved when no additional effect included. When including the edge bond relaxation and third nearest neighbor, disagreement is pointed between real and mode results, which is due to the approximation when transform the real space into uncoupled mode space.

Table 1 Error range of current calculation and total simulation time for real and mode space approaches. Row 1 shows the relative error range for current calculation between real and mode space approach, when supply voltage is 0.5V. Last two rows are the simulation time (not including Poisson equation solving time) when $V_G$=0.5V and $V_D$ increasing from 0 to 0.5V by 0.1V per step. The comparison shows that the mode space approach reduces the computational cost by about one order of magnitude (the simulations were run on a single 3.0GHz CPU).

| | Model 1 | Model 2 | Model 3 |
|---|---|---|---|
| Min. ~ Max. Error($V_{DD}$=0.5V) | 0.07~0.39% | 2.45~2.88% | 4.62~5.52% |
| Simulation time for Real space | 2554s | 3079s | 3908s |
| Simulation time for Mode space | 101s | 101s | 490s |

Table 1

FIGURES

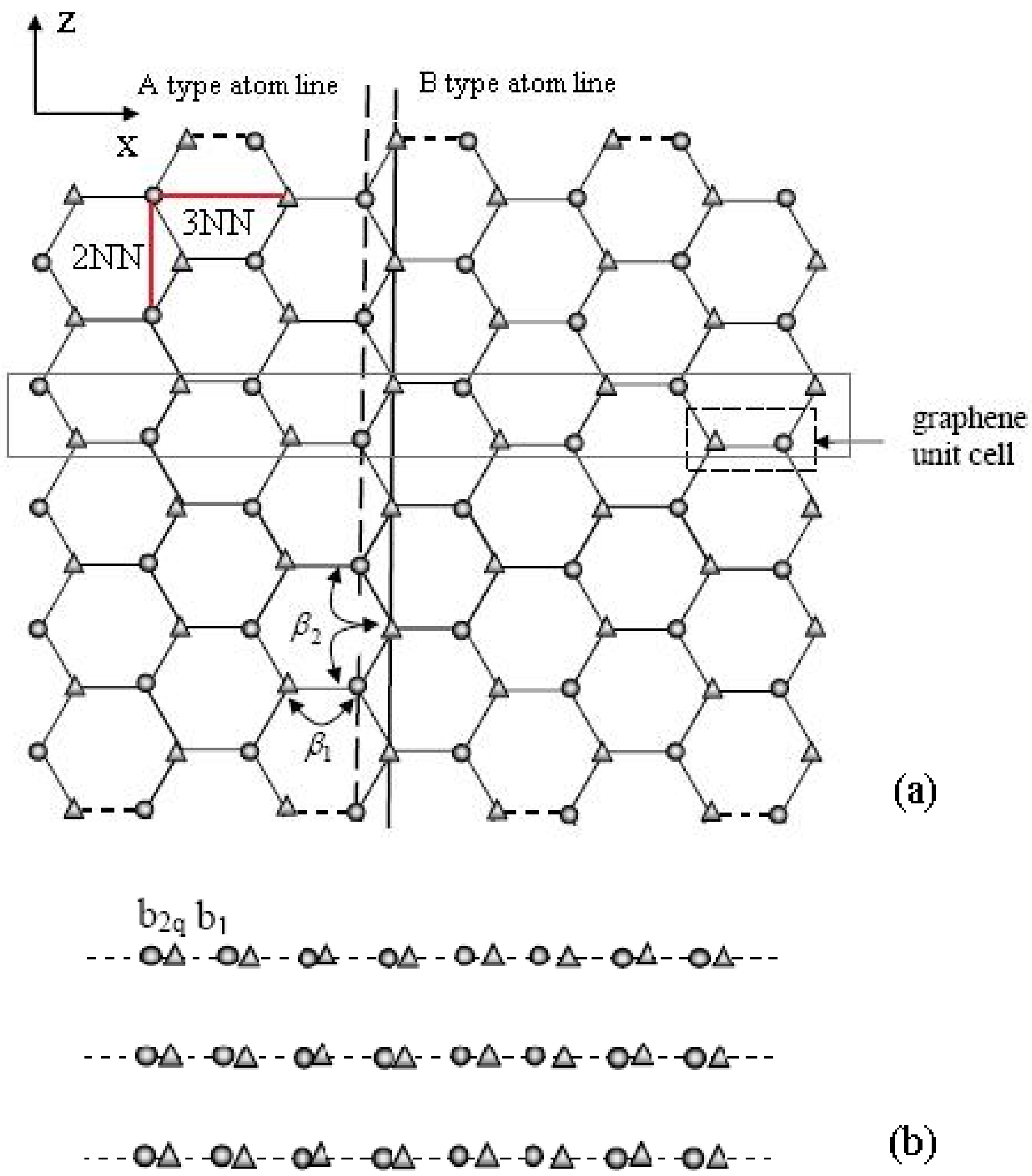

Fig. 1

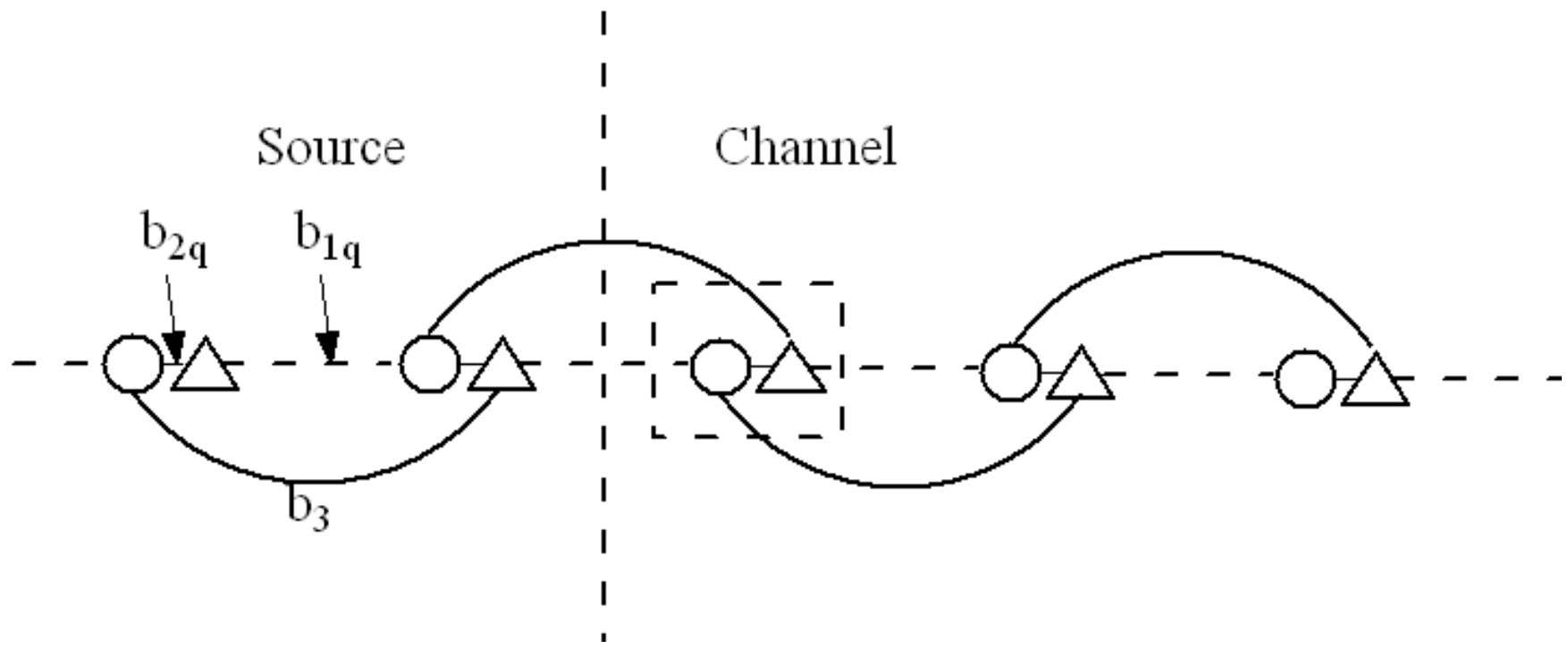

Fig. 2

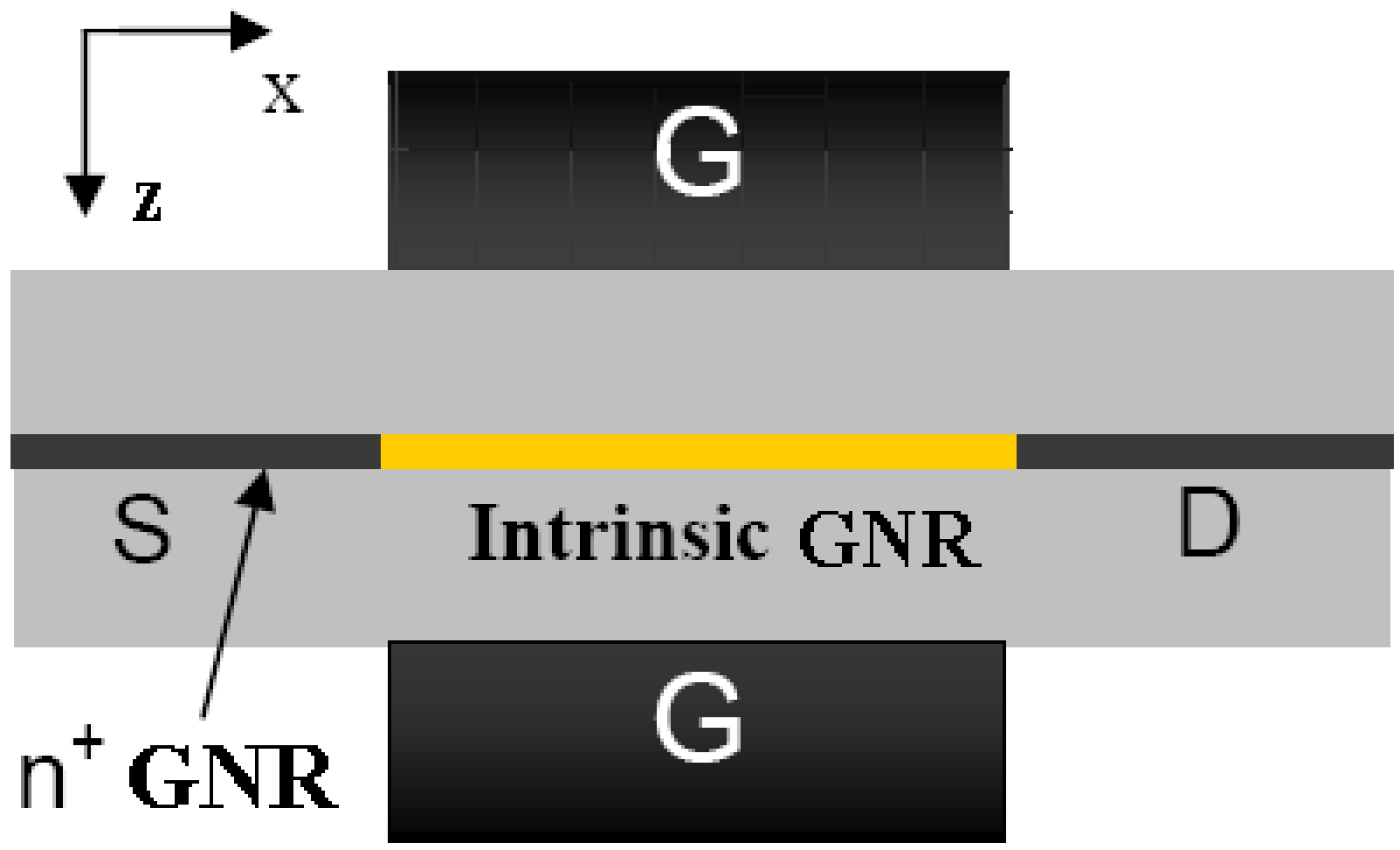

Fig. 3

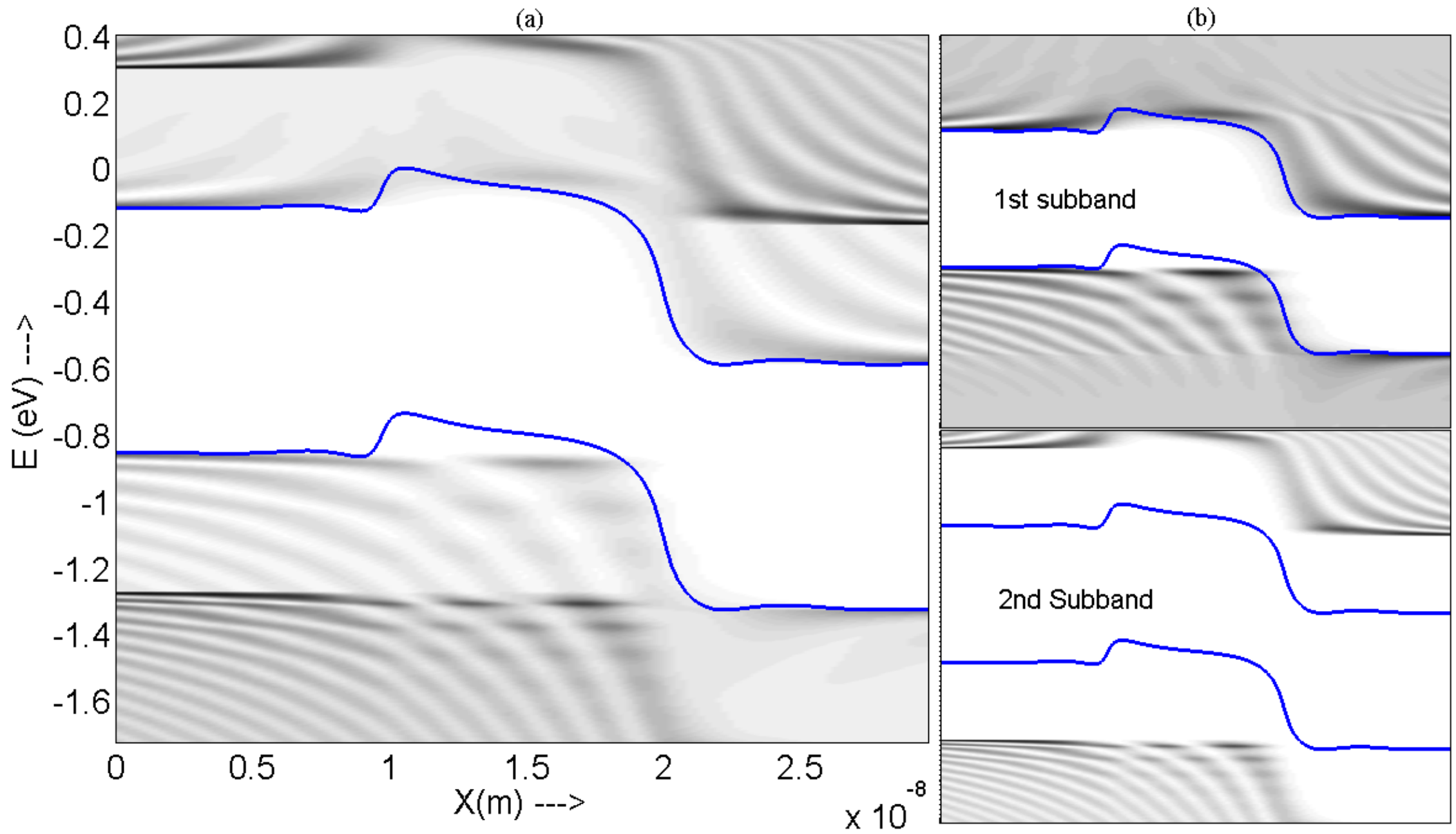
(a)
(b)
0.4
0.2
0
-0.2
-0.4
-0.6
-0.8
-1
-1.2
-1.4
-1.6
E (eV) --->
0
0.5
1
1.5
2
2.5
X(m) --->
x 10^-8
1st subband
2nd Subband

Fig. 4

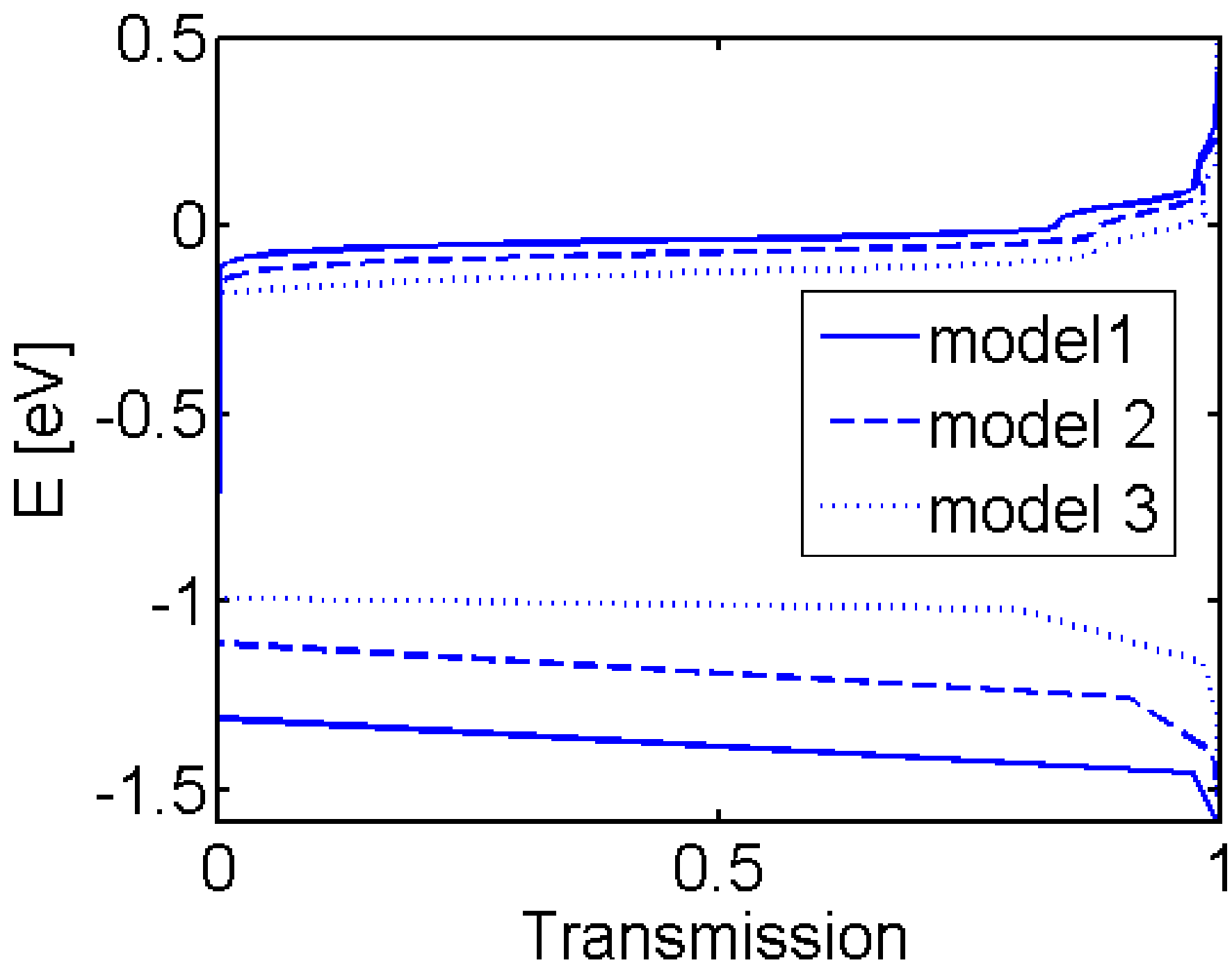

Fig. 5

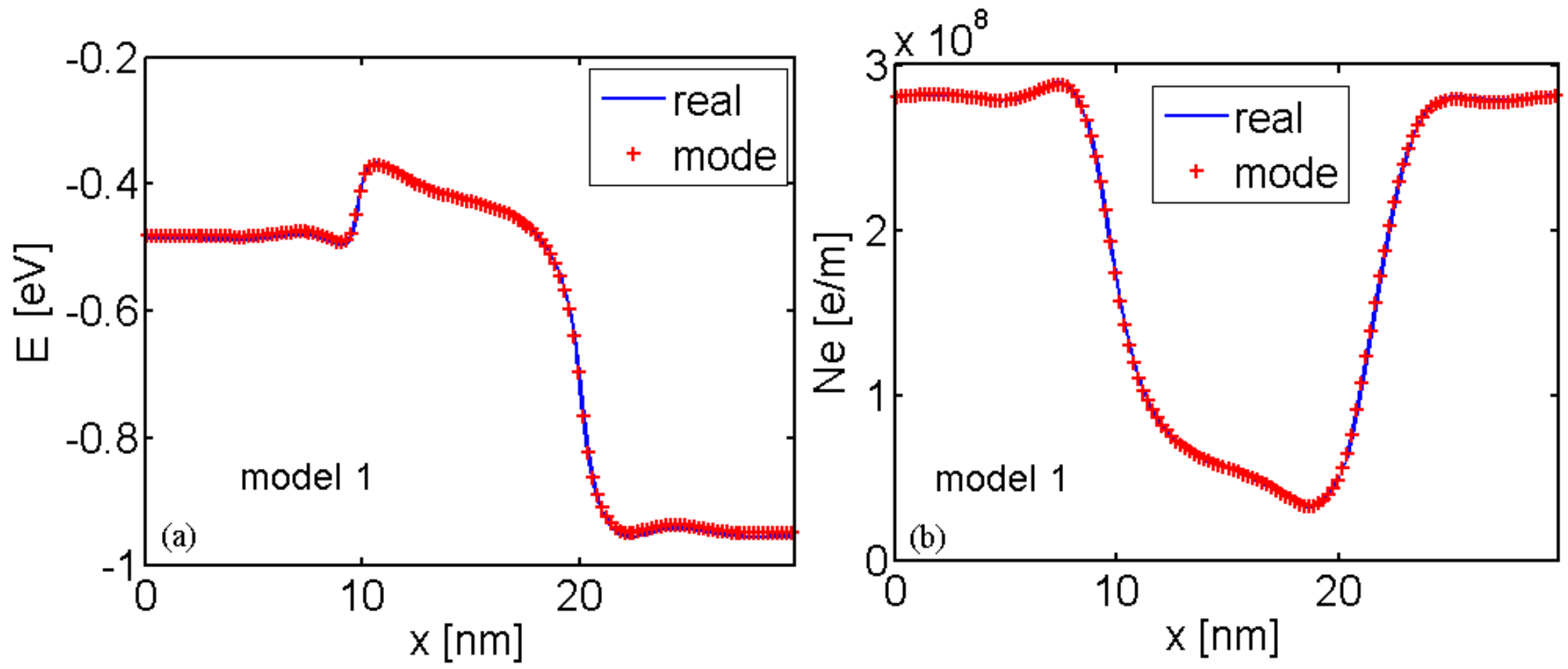
-0.2
-0.4
-0.6
-0.8
-1
E [eV]
real
mode
model 1
(a)
0
10
20
x [nm]
3
x 10^8
2
1
0
Ne [e/m]
real
mode
model 1
(b)
0
10
20
x [nm]

Fig. 6

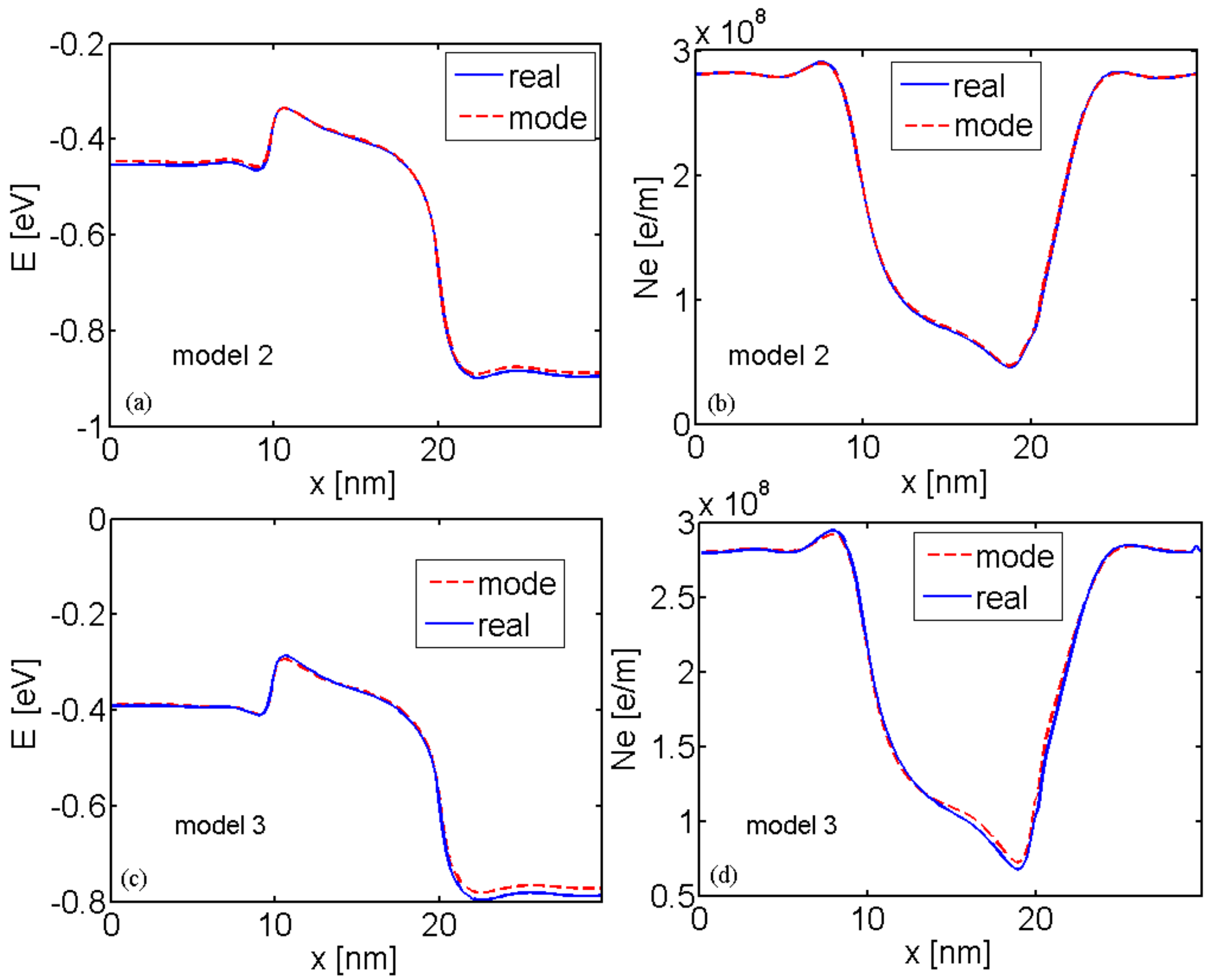

(a)
model 2
real
mode
E [eV]
x [nm]
(b)
model 2
real
mode
Ne [e/m]
x 10^8
x [nm]
(c)
model 3
mode
real
E [eV]
x [nm]
(d)
model 3
mode
real
Ne [e/m]
x 10^8
x [nm]

Fig. 7

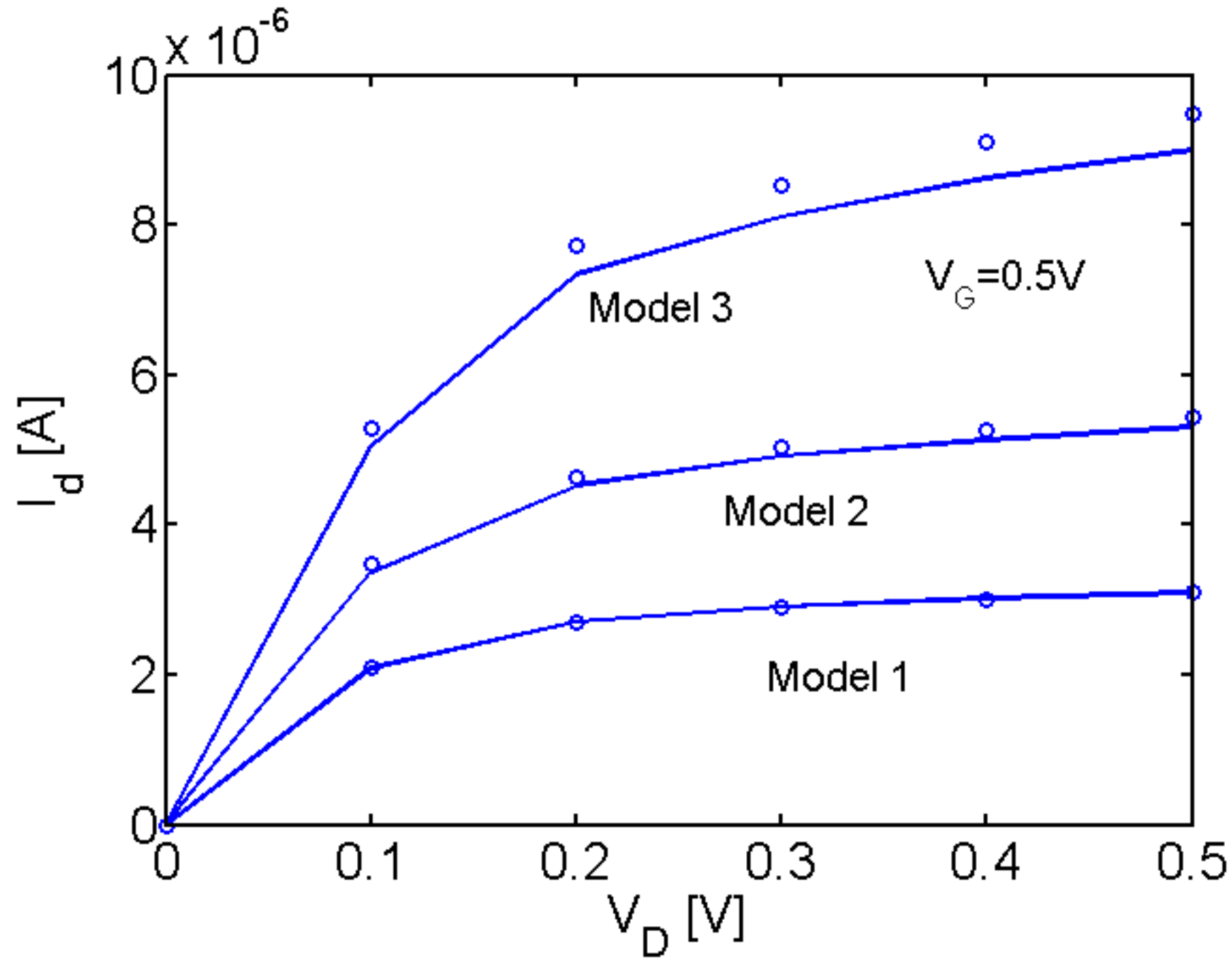
x 10^-6
10
8
6
4
2
0
I_d [A]
0
0.1
0.2
0.3
0.4
0.5
V_D [V]
Model 3
V_G=0.5V
Model 2
Model 1

Fig. 8